# Mixed regime of light-matter interaction revealed by phase sensitive measurements of the dynamical Franz-Keldysh effect

Fabio Novelli[1], Daniele Fausti*[1,2], Francesca Giusti[1], Fulvio Parmigiani[1,2], Matthias Hoffmann[3].

(1) Department of Physics, Università degli Studi di Trieste, 34127 Trieste, Italy

(2) Sincrotrone Trieste S.C.p.A., 34127 Basovizza, Italy

(3) SLAC National Accelerator Laboratory Menlo Park, California, United States

* Correspondence to: daniele.fausti@elettra.trieste.it

**The speed of ultra-fast optical switches is generally limited by the intrinsic electronic response time of the material. Here we show that the phase content of selected electromagnetic pulses can be used to measure the timescales characteristic for the different regimes of matter-light interactions. By means of combined single cycle THz pumps and broadband optical probes, we explore the field-induced opacity in GaAs (the Franz-Keldysh effect). Our phase-resolved measurements allow to identify a novel quasi-static regime of saturation where memory effects are of relevance.**

The intrinsic charge carrier dynamics limits the speed at which material properties can be manipulated by an electromagnetic (EM) pulse [1,2,3]. When the characteristic electronic timescale is comparable to the duration of a single cycle of the EM field, the details of the field carrier-envelope phase determine the material response. Under these conditions matter can be excited into the non-perturbative regime [4,5,6,7] and a full quantum mechanical treatment is needed. An important quantity describing the "strength" of the field-matter interaction is the ponderomotive energy [8,9]

$$U_P = \frac{e^2 \mathrm{E_0}^2}{4m\Omega^2} = \hbar\Omega \cdot \gamma, \tag{1}$$

which is defined as the mean kinetic energy of a particle of mass $m$ and charge $e$ which oscillates in the ac-electric field $E(t) = \mathrm{E_0}\cos(\Omega t)$. For $\gamma \ll 1$ the matter-field interactions can be described perturbatively while for $\gamma \gg 1$ the field strength is much larger than the photon energy and a full treatment of the electromagnetic field is needed.

In this paper we report a study of the dynamical Franz Keldysh effect (DFKE) in the transition region between those two limits ($1 < \gamma < 30$). Single-cycle THz electromagnetic pulses with maximum field amplitudes between ~30 kV/cm ($\gamma \sim 1$) and 100 kV/cm ($\gamma \sim 30$) are used to gate the transmission in bulk gallium arsenide. The changes in transmission are dramatic and equal 60% transmission loss through a 0.4 mm sample. Our amplitude and phase-dependent study of the THz-induced opacity in bulk GaAs allows for the identification of a novel anomalous regime of the Franz Keldysh effect (FKE), that has a static intensity dependence but a highly non-trivial phase evolution.

The static Franz Keldysh effect can be described as follows: in a semiconductor, the relative motion of an electron-hole pair in a uniform and static electric field $F$ is described by a one-dimensional and time-independent Schrödinger equation,

$$\left(\frac{\hbar^2}{2\mu}\frac{\partial^2}{\partial z^2} + |e|Fz + E_z\right)\varphi(z) = 0, \qquad (2)$$

whose solution is an Airy function [10,11,12,13] $\varphi(z) \rightarrow \varphi(\xi) \propto Airy(\xi)$ obtained by the substitution $z \rightarrow \xi = -\frac{E_z}{\sigma(F)} - z\left(\frac{2\mu|e|F}{\hbar^2}\right)^{1/3}$, with $\sigma(F) = \left(\frac{e^2\hbar^2F^2}{2\mu}\right)^{1/3}$ the "electro-optical energy" of the field. The Airy function contains the salient features of the static FKE: it has an exponential tail for small and negative values of the argument (that can be related to below-gap field-induced absorption), and it approaches a plane-wave like solution for positive and increasing arguments (implying the above-gap oscillations).

Figure 1a shows the absorption of a conventional semiconductor in a static and uniform electric field. The standard "square root" absorption (black curve) in the presence of an electric field (red curve) is strongly perturbed in the energy region across the band-gap $E_{Gap}$[14,15]: an exponential-tail absorption appears below $E_{Gap}$ (also known as electroabsorption) and an oscillatory behaviour in frequency of the optical properties of the semiconductor is revealed above the energy of the gap.

When the static electric field is replaced by a time dependent one (Fig.1b) the response of the system is described by the Dynamical Franz-Keldysh effect (DFKE) [16,17]. The DFKE is qualitatively similar to the static FKE below the gap (i.e. both effects exhibit an exponential absorption tail[18]) but the above-gap oscillations are much weaker and the absorption edge is blue-shifted by the ponderomotive energy $U_P$ [9,19]. When $U_P$ is of the same order of magnitude of the photon energy ($\gamma \sim 1$) the conduction and valence bands cannot follow in a quasi-static way the applied EM field and the field-induced opacity is better described by the DFKE. On the other hand, for growing $U_P$ ($\gamma \gg 1$) the effects become better and better described by a quasi-static FKE, which is the proper model for a uniform dc field ($\gamma = \infty$).

**Results**

Field induced optical absorption experiments

Here we study the FKE at the transition between a dynamical ($\gamma \sim 1$) and a quasi-static regime ($\gamma = 30 \gg 1$). To this purpose we developed a pump-probe experiment which uses strong almost single cycle EM pulses at THz frequencies as pump and broadband near-infrared pulses as probe. It is important to note that the probe pulses used in our experiments are shorter than the THz wavelengths so that our technique allows the phase sensitive measurement of the DFKE.

Fig.2 shows the THz electric field in the time domain (a) and a spectrogram of the observed transmission change $\Delta T(t)/T$, with $\Delta T(t)$ is the time-dependent perturbed transmission and $T$ is the static transmission in GaAs (b). At the peak of the THz field the transmission is reduced by up to $60\%$ (Fig.2b) at photon energies just below the band gap. This large modulation is caused by the strong sub-gap absorption as shown for the static FKE in Fig.1(a).

Fig.2c shows the measured $\Delta T(t=0 \text{ ps})/T$ as function of photon energy (red curve) compared to the transmission change calculated from eq.2 [13] assuming a static electric field of $100\,\mathrm{kV/cm}$ (black curve) and a square-root gap. The only free parameter of the calculation is a phenomenological scaling factor accounting for the size of the matrix elements of the dipole transitions. It should be noted that taking into account realistic shapes of the equilibrium absorption (including excitonic effects), rather than a simple square-root gap, gives only minor differences in the field driven absorption.

While the shape of the change as function of photon energy is accurately described at every time step, a comparison between Fig.2a and Fig.2b reveals that we cannot reproduce the observed *temporal dependence* of the transient transmission by calculating the static FKE with the THz field profile shown in Fig 2a. Nevertheless, this suggest that an effective electric field can describe the temporal evolution of the transients, as will be discussed in the following.

### THz pump reflectivity probe

In order to study the transition region between the static and dynamical FKE, and to elucidate the phase evolution of the field-induced variations of the optical properties, we carried out single-color reflectivity measurements for different pump intensities ($1 \leq \gamma \leq 30$). Reflection geometry experiments increase temporal resolution by avoiding dephasing of pump and probe pulses due to the slightly different group velocities in a bulk sample.

Fig.3b shows the reflectivity change $\Delta R(t)/R$ at 900 nm induced by the presence of the strong field in Fig.3a. The shape of the transient reflectivity changes dramatically as a function of the pump field strength (Fig.3b). In order to highlight the main result of this report, i.e. the different phase content of the response to fields with different intensities, in Fig.3c we plot the normalized $\Delta R(t)/R$ at low and high pump intensity together with the modulus of the THz field (dashed line). Fig.3d shows the maximum $\Delta R/R$ versus THz field strength. We observe an almost linear dependence for $\Delta R/R$ up to $I \approx 50\%$, while a sub-linear behaviour appears at higher field strengths. From the theory of the static FKE we expect that the pump-induced opacity scales with the amplitude of the applied electric field [20], while the DFKE predicts a linear dependence with pump intensity [18]. Hence, from the maximum amplitude of the transient reflectivity (Fig.3b), we can identify a dynamical Franz-Keldysh regime for low THz fields and an anomalous behaviour for field strengths higher than 70 kV/cm. The analysis of the transient reflectance in this regime reveals a static-like dependence.

**Discussion**

As indicated earlier, a static FKE model can describe the transient transmission at all wavelengths at each time delay (Fig.2c) once a parameterized electric field is provided. This evidence lead us in the search of a phenomenological expression for an effective ac field that, once used within the static FKE theoretical framework, reproduces the fluence and phase dependent evolution of the transient reflectivity reported in Fig.3b.

In order to describe the anomalous regime investigated (Fig. 3b, $I < 50\%$) we follow an heuristic approach. When the potential in eq.1 is a function of time, the corresponding one-dimensional time-dependent Schrödinger equation [21,22,23], with a potential that is linear in the space variable, has a solution known as Airy packet [23]. The Airy packet is similar to the Airy function, which is the solution to the static case, but its argument depends on the integral of the potential over time.

By inspection of the blue curve in Fig.3c ($\gamma \sim 7$) we notice that for low intensity the peak in the THz field at t=-0.6 ps (dashed line) has no effect on the transient reflectivity. Moreover, at later times (t=+0.7 ps) the second peak in the field give rise, slightly shifted, to a large variation of the reflectivity. The ratio of the THz field at t=+0.7 ps over the one at t=0 ps is much smaller than the ratio of the two highest variations in reflectivity $\Delta R(t = 0.9\ \text{ps})/\Delta R(t = 0.1\ \text{ps})$. This indicates that, for low pump intensities, the optical response at time t has *memory* of the THz field at all previous times or, equivalently, that an integral function of the potential is governing the transient optical properties. This behaviour is consistent with the solution of Berry [23] and with the expectation in the dynamical regime of the Franz-Keldysh effect.

The phase dependence of the transient reflectivity is dramatically modified upon increasing the strength of the ac field (red curve in Fig.3a). Apart from the expected broadening at saturation of the optical response [24], it is evident that the three peaks of the applied THz field at t=-0.6 ps, t=0 ps, and t=0.7 ps, are responsible for three corresponding structures in $\Delta R(t)/R$ at the same pump-probe delays (compare red and dashed curves in Fig.3c). This is in qualitative agreement with the static models for the FKE predicting a response proportional to the electric field amplitude. Nevertheless, the details of the phase-evolution are surprising. The transient reflectivity due to the first THz semiperiod (t=-0.6 ps) is much higher with respect to the reflectivity changes due to the central THz peak. These observations highlight a new regime for the FKE where the dynamic response saturates and the phase dependence approaches a quasi-static regime.

Led by all the previous consideration we could formulate a phenomenological expression for the effective ac field $F_{eff}(p,t)$ to be used within a dc FK model:

$$F_{eff}(p,t) = \alpha(p)\left(|F(t)|\frac{A(I^2(t))}{1+B(I^4(t))} * g(p,t)\right),$$  (4)

where $F(t)$ is the THz field amplitude (which is an experimental input measured by electro-optical sampling), $I(t) = F^2(t)$ is the time-dependent intensity, $A$ and $B$ are integral functions of their respective arguments, $p$ is the normalized peak intensity, $\alpha(p)$ a phenomenological analytical function of $p$, and $g(p,t)$ a convoluted Gaussian. The explicit form of eq.4 can be found in the supplemental material. In Fig.4a we plot the measured $\Delta R(t)/R$ and the results of the simulation with the effective field given in eq.4 for different pump intensities. Apart from the slow dynamics observed at delays higher than 2 ps, that are related to incoherent pump-induced heating effects not included in eq.4, our model contains the main features of the temporal and intensity dependence of the transient reflectivity probed at 900 nm. At low intensities the effective field is dominated by the contribution $A(I^2(t))$ which is consistent with the DFKE regime where the solution of the time-dependent Schrödinger equation depends on the integral of the potential over time [23]. On the other hand, when the intensity is increased, $F_{eff}$ is suppressed by the renormalization due to higher order corrections at the denominator, leading to the recovery of the sub-linear dependence in the field intensity characteristic of the static FKE. The anomalous phase-dependence of the FKE with strong pump originates in our model from the integral of the higher order power of the intensity at the denominator, indicating that the optical response of GaAs at time t still carries a memory of the history of the applied electric field in the quasi-dc limit.

In conclusion, we performed a systematic study of the transition region between the dynamical and quasi-static regimes of matter-light interaction by detecting the ac field induced opacity in bulk GaAs. The analysis of the dependence of the reflectivity at 900 nm on the EM field amplitude and phase revealed the existence of a novel regime of saturation for the dynamical Franz-Keldysh effect. Our phase-resolved technique allows to observe directly the memory effects and to establish a phenomenological model that uses the static FKE solution combined with an effective field to describe the new regime. We demonstrate, by tuning the amplitude of single cycle pump pulses, that the temporal response of the optical properties of GaAs exhibits a highly non trivial phase content. Finally, we suggest that this novel regime will be of relevance for ultra-fast optical gating devices.

**Methods**

The single-cycle THz pulses, with energies of 2 µJ and field strengths exceeding 100 kV/cm, were generated by optical rectification in $LiNbO_3$ using the tilted pulse front technique [25]. The pulses were characterized by electro-optical sampling using a 150 µm GaP crystal, revealing their field strength and spectral content which is peaked at 0.8 THz. The total pulse energy was measured independently with a calibrated pyroelectric detector. A pair of wiregrid polarizers was used for controlled attenuation of the THz field without changing its temporal profile.

We can estimate from eq.1, in the free electron approximation and for central frequency $\Omega \approx 1\,THz$, that $\gamma$ for the EM pulses used in the experiments range from $\gamma \approx 1$ for maximum field strength ~30 kV/cm up to $\gamma \approx 30$ for the pulses with the highest energy (~100 kV/cm). As reported in Fig.1b, with this field parameters we expect to cross the boundaries between the DFKE and a quasi-static regime upon increasing the field amplitude.

The femtosecond laser pulse was split off to generate white light in a 2 mm thick sapphire plate. The white light was focused onto the sample using a thin $CaF_2$ lens. The chirp of the probe continuum in the near IR region of interest is estimated to be less than 100 fs. A fiber-coupled spectrometer was used to measure the time dependent transmission change ΔT(t)/T. We used a lightly doped n-type GaAs ($8x10^{15}$ $cm^{-3}$) in order to suppress THz reflections within the sample that would lead to echo artifacts in the time domain data. Additionally we carried out single-color pump-probe measurements in reflectivity with bandpass filters (10 nm full width half maximum) between the sample and photodetectors.

**Aknowledgements**

The authors are grateful to Andrea Cavalleri as part of the measurements presented here were performed in the Max Plank Institute for Structural Dynamics, Hamburg, DE. We would also like to acknowledge Dr. Marco Malvestuto, Dr. Claudio Giannetti, Dr. Federico Cilento and P. H. M. van Loosdrecht for critical review of the manuscript. This work has been supported by the European Union, Seventh Framework Programme, under the project GO FAST, grant agreement no. 280555.

**Author contribution statement**

FN, DF performed the analysis and wrote the manuscript, MH performed the experiments and participated in the writing process. FG contributed to the experiments and FP to the discussions.

**Competing financial interests**

The authors declare no competing financial interests.

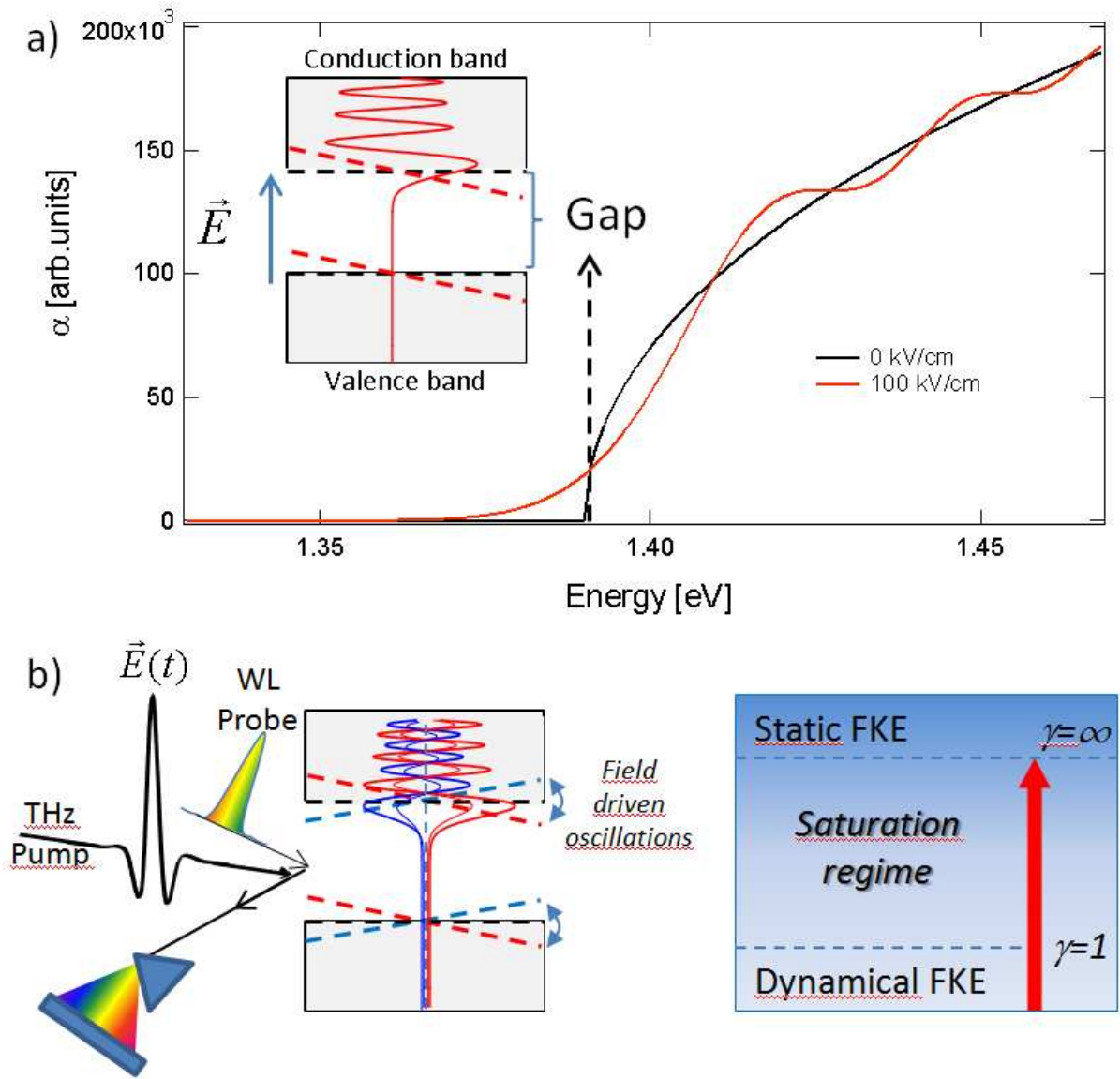

Figure 1: **The Franz–Keldysh effect**. a) The unperturbed absorption edge of a semiconductor (black curve) displays, in a uniform electric field, below-gap absorption and above-gap oscillations (red curve). Insert: a static electric field tilts the bands enhancing sub-gap tunnelling leading to below-gap absorption. b) Dynamical Franz-Keldysh effect detected by pump-probe THz spectroscopy. The bands are titled with a non-trivial phase relation with the applied ac fields. Right: sketch of the two limits of the FKE, that is static for infinite $\gamma$ and dynamic for $\gamma$ close to 1 (see text).

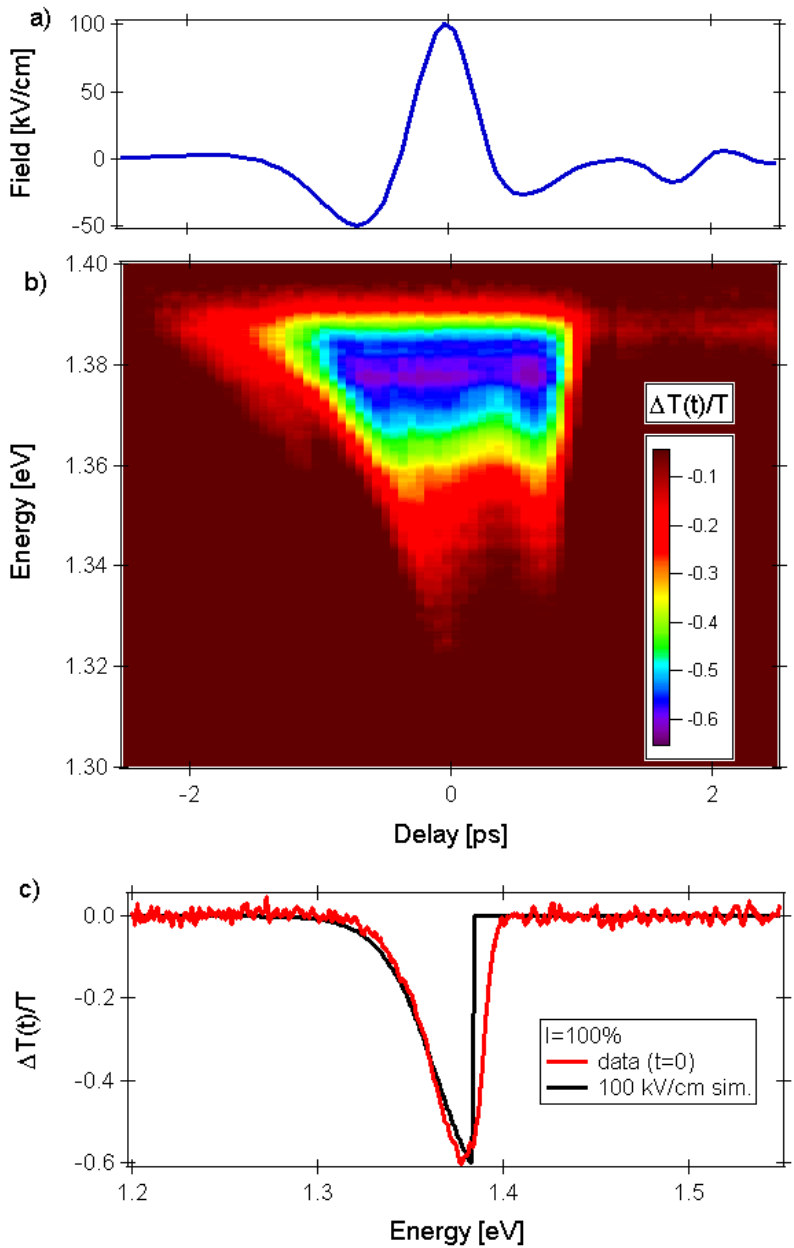

Figure 2: **THz-driven Franz-Keldysh effect**. (a) THz field detected by electro-optical sampling. (b) Time-resolved variation of the transmission in GaAs as a function of pump-probe delay and probed energy. (c) Wavelength-dependent Franz-Keldysh effect. Vertical section at t=0 (red curve) and simulation based on the static FK effect with an applied field of 100 kV/cm (black curve). Note that (a) and (b) have the same abscissa.

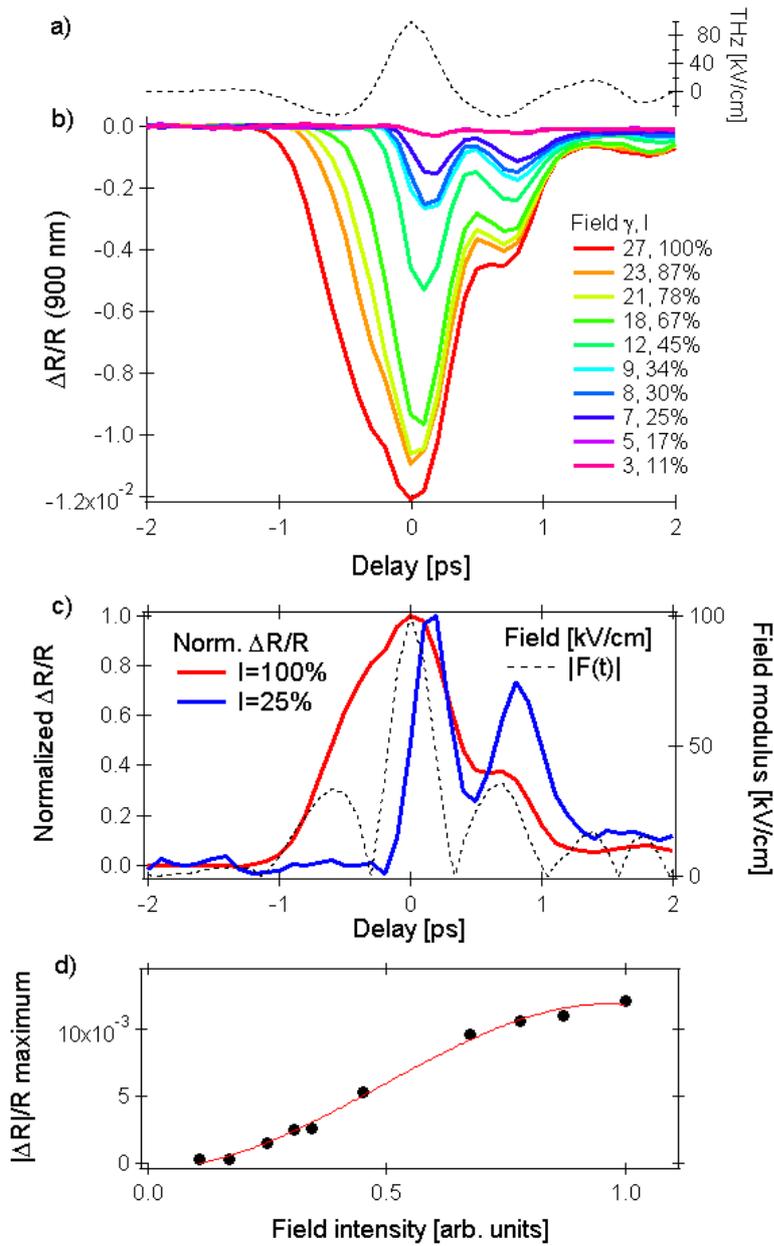

Figure 3. **THz driven reflectivity changes**. a) Pump electric field temporal profile measured by electro optical sampling. b) Transient reflectance induced by electric field of different intensity ($\gamma$ is the Keldysh parameter, see text for details). c) Normalized relative variation of the reflectivity at 900 nm (1.38 eV) for highest (red) and lowest (blue) pump intensities. The absolute value of the THz field, whose phase content is independent of the intensity of the pump, is reported for comparison (dashed black curve). d) Maximum value of $\Delta R(\omega,t)/R_0(\omega)$ as a function of pump amplitude. The red line is a guide for the eyes.

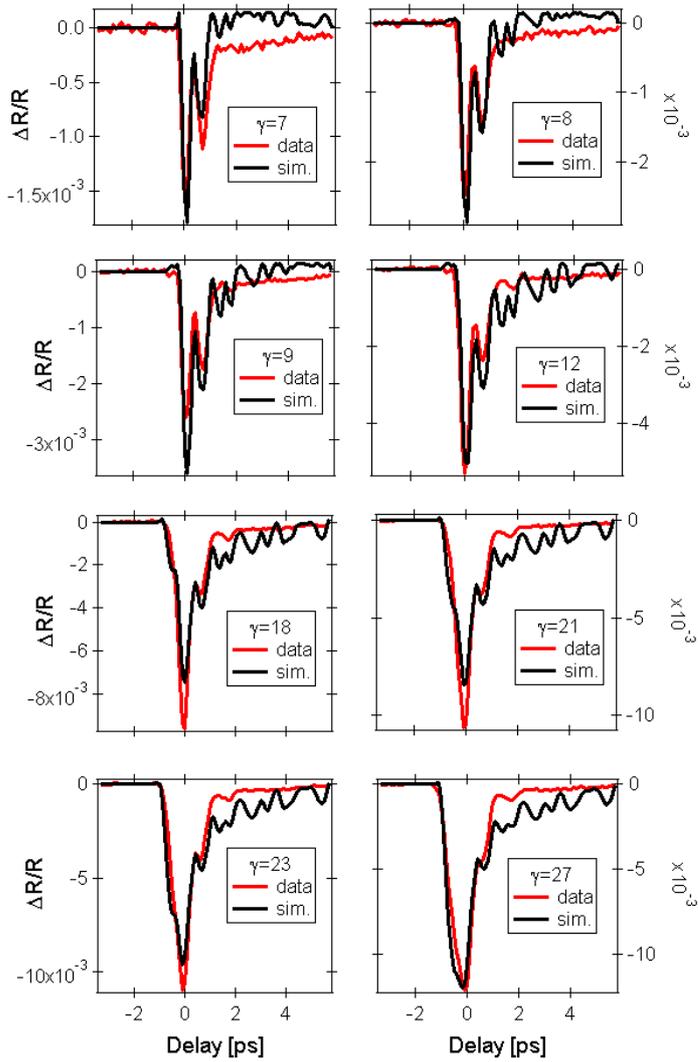

Figure 4. **Transition from dynamical to quasi-static FKE.** a) $\Delta R(\omega, t)/R_0(\omega)$ at 900 nm as a function of $\gamma$ and time delay, data (red) and model (black).